\documentstyle[amssymb,aps,prl,multicol,epsfig]{revtex}
\begin{document}
\draft  \title
{Response to tilted magnetic fields in $Bi_2Sr_2CaCu_2O_8$ with columnar defects : 
Evidence for transverse Meissner effect.}
\author{V. Ta Phuoc, E. Olive, R. De Sousa, A. Ruyter, L. Ammor and J.C. Soret
}
\address{Laboratoire d'Electrodynamique des Mat\'eriaux Avanc\'es 
 Universit\'e François Rabelais - CNRS FRE 2077 - CEA LRC M01
  Parc de Grandmont- 37200 Tours - France
}
\maketitle

\begin{abstract}
	The transverse Meissner effect (TME) in the highly layered superconductor $Bi_2Sr_2CaCu_2O_{8+y}$ with 
	columnar defects is investigated by transport measurements. We present detailed evidence for the 
	persistence of the Bose-glass phase when $H$ is tilted at an angle $\theta < \theta_c \left( T \right)$ 
	away from the column direction: (i) the variable-range vortex hopping process for low currents crosses 
	over to the half-loops regime for high currents; (ii) in both regimes near $\theta_c(T)$  the energy 
	barriers vanish linearly with $\tan\theta$ ; (iii) the transition temperature is governed by 
	$T_{BG}(0) -T_{BG}(\theta) \sim |\tan\theta| ^{1/\nu_{\perp}}$ with $\nu_{\perp}=1.0 \pm 0.1$. 
	Furthermore, above the transition as $\theta\rightarrow\theta_c^+$, moving kink chains consistent with 
	a commensurate-incommensurate transition scenario are observed. These results thereby clearly show 
	the existence of the TME  for $\theta < \theta_c(T)$ .
 \end{abstract}

\begin{multicols}{2}
  \narrowtext
  
 	The mixed state in high-$T_C$ superconductors has been of considerable interest over the past ten 
	years. The reason is that combined effects between temperature, elasticity, intervortex interactions, 
	and disorder yield a large variety of new thermodynamic vortex phases \cite{1}. In particular, the 
	theory of vortex pinning by correlated disorder such as twin planes or artificial columnar defects has 
	been considered by Nelson and Vinokur (NV) \cite{2} and Hwa {\em et~al} \cite{3}. 
	In the case of parallel columns, NV have shown that if the applied magnetic field {\bf H} is aligned with the columnar 
	defects, the low temperature physics is similar to that of the Bose glass (BG) \cite{4}, with the flux lines strongly 
	localized in the tracks leading to zero resistivity. When {\bf H} is tilted at an angle $\theta$ away from the column 
	direction, the BG phase with perfect alignment of the internal flux density {\bf B} parallel to the columns is predicted 
	to be stable up to a critical transverse field $H_{\perp c}$ producing the so-called transverse Meissner effect (TME) 
	\cite{2,5}. For $\theta >  \theta_c \equiv \arctan(\mu_0H_{\perp c}/B)$, the linear resistivity is finite resulting from 
	the appearance of kink chains along the transverse direction as discussed in Ref.\ \onlinecite{6}. Finally, above a still 
	larger angle, the kinked vortex structures disappear and {\bf B} becomes collinear with {\bf H}. Similar scenarios have 
	also been developed for pinning of vortices by the twin boundaries \cite{2} and by the layered structure of the coumpound 
	itself \cite{7}.\\
		\indent Recently, the TME in untwinned single crystals of $YBa_2Cu_3O_7$ (YBCO) with columnar defects, has been 
	observed using an array of Hall sensors \cite{8}. Previous magnetization or magnetic-torque measurements in anisotropic 
three-dimensional superconductors, such as YBCO, gave support to the presence of vortex lock-in phenomena, due to pinning 
by the twin boundaries \cite{9} or by the interlayers between the Cu-O planes \cite{10}. In the case of highly layered 
superconductors, such as $Bi_2Sr_2CaCu_2O_{8+y}$ (BSCCO), although the lock-in transition was observed for {\bf H} tilted 
away from the layers \cite{11}, the existence of the TME due to the pinning by columnar defects remained an open question.\\
	\indent In this letter, we present measurements of the electrical properties near the glass-liquid transition in BSCCO 
single crystals with parallel columnar defects, as a function of $T$ and $\theta$ for filling factors $f< 1$. We find that 
when $\theta$ is less than a critical angle $\theta_c(T)$, the resistivity $\rho (J)$  vanishes for low currents. A 
detailed {\it quantitative} analysis of  $\rho (J)$ indicates that the creep proceeds via variable-range vortex hopping 
(VRH) at low currents due to some on-site disorder \cite{12}, crossing over to the half-loop regime at high currents. 
For $\theta > \theta_c(T) $, the very signature of a kinked vortex structure, consistent with a commensurate-incommensurate 
(CIC) transition scenario in (1+1) dimensions \cite{6}, is deduced from the measurement of the critical behavior of the 
linear resistivity. All these results clearly demonstrate that when the field is tilted at $\theta < \theta_c$ away 
from the defects, the flux lines remain localized on columnar defects, and hence, the BG exhibits a TME.\\
	\indent The BSSCO single crystal was grown by a self-flux technique, as described elsewhere \cite{13}. The crystal of 
$1\times 1\times 0.030\ mm^3$ size with the c-axis along the shortest dimension has a $T_C$ of $89\ K$, a transition 
width of $\sim 1\ K$, and was irradiated along its c-axis with $5.8\ GeV$ Pb ions to a dose corresponding to 
$B_{\Phi}=1.5\ T$ at the Grand Acc\'el\'erateur National d'Ions Lourds (Caen, France). The crystal was mounted onto a 
rotatable sample holder with an angular resolution better than $0.1^{\circ}$ in a cryostat with a $9\ T$ magnet. 
Isothermal $I-V$ curves were recorded using a $dc$ four-probe method with a sensitivity of $\sim 10^{-10}\ V$ and a 
temperature stability better than $5\ mK$. {\bf H} was aligned with the tracks using the well-known dip feature 
occurring in dissipation process for $\theta =0^{\circ}$, and was tilted away from the column direction for 
$f\equiv \mu_0H_{\parallel}/B_{\Phi}$ fixed. We present in this paper the data obtained for $f=1/3$.\\
	\indent Fig. (1) shows a typical log-log plot of $V/I-I$ curves obtained varying $\theta$ for $T< T_{BG}(0)$ fixed. 
We observe a well-defined angular crossover at an angle $\theta_c$. For low angles $(\theta< \theta_c)$, 
$V/I-I$ data suggest a glassy vortex-like state resulting from the persistence of the BG phase previously observed 
in this crystal for $\theta =0^{\circ}$ \cite{12}. On the contrary, upper curves $(\theta> \theta_c)$ display an 
Ohmic regime, and hence, indicate a liquid vortex-like state.\\ 
	\indent We first focus on the angular range  $(\theta< \theta_c)$ where a divergent conductivity with vanishing current is 
observed, and we consider the scenario where the BG phase is stable. Thus, one expects that excitations of some localized 
vortex lines lead to a nonlinear resistivity given by \cite{2}:
   \begin{eqnarray}  
\rho(J)=\rho_0\ exp\big(-\tilde E_K(J_c/J)^{\mu}/k_BT\big)
   \end{eqnarray}
where $\rho_0$  is a characteristic flux-flow resistivity and $\tilde E_K(J_c/J)^{\mu}$ represents the barriers 
against vortex motion. This expression is predicted to hold for various regimes of different behavior as the current 
density probes different length scales in the BG. Here, $\tilde E_K$ acts as a scaling parameter for the pinning energy 
and $J_c$ is the characteristic current scale of the creep process. When the current density is large enough that the 
growth of vortex-loops excitations of a line from its pinning track does not reach out the neighboring 
tracks, the half-loops excitations are relevant and lead to an exponent $\mu=1$ and $J_c\equiv J_1=U/(\Phi_0d)$, 
where $U$ is the mean pinning potential and $d=\sqrt{\Phi _0/B_{\Phi }}$. With decreasing current density, critical 
size of half-loops increases with the result that some disorder in the pinning potential becomes relevant. This 
situation yields the VRH process characterized by $\mu=1/3$ and by another important current scale 
$J_c\equiv J_0=1/(\Phi _0g(\tilde\mu)d^3)$, where $g(\tilde\mu)$  denotes the density of pinning energies at the 
chemical potential of the vortex system. One expects that both $J_1$ and $J_0$ are insensitive to $H_{\perp}$. 
We therefore consider that the only important effect of $H_{\perp}$ is to lower the scaling parameter for 
the pinning energy in Eq.\ (1), according to the formula :
   \begin{eqnarray}  
\tilde E_K=E_K-\epsilon ^2\Phi _0dH_{\perp}
   \end{eqnarray} 
where $E_k$ is the mean kink energy, and the energy gain due to the tilt is obtained from the isotropic result 
$\Phi _0dH_{\perp}$  by applying the scaling rule Eq.\ (3.12) of the Ref.\ \onlinecite{1}.\\ 
	\indent We now present the following two-step method of fitting Eqs.\ (1) and (2) to the data, which allows to get a good 
understanding of the physics in our experiment. First, we plot in Fig.\ 2 the natural logarithm of V/I versus 
$\tan\theta$  for $I$ fixed. Note that $\tan\theta$  happens to be here directly proportional to $H_{\perp}$ since $f$ 
is maintained constant. A linear variation $\ln(V/I)=A+B \tan\theta$ (expression designated 
below as $E1$) with $A$ and $B$ being current-dependent parameters is found for $\theta$ varying up to $\theta_c$ 
(solid lines), whereas above $\theta_c$ data deviate from this behavior (dotted lines). This finding is another argument 
supporting a true angular transition at $\theta_c$, as suggested in Fig.\ 1. Here, it should be noted that we draw such 
a conclusion from the observation of two {\it independent} regimes of {\it different} behavior. One leads to a vanishing 
linear resistivity as $\theta\rightarrow\theta_c^+$, and the other is evidenced by probing the regions of 
nonlinear resistivity above and below $\theta_c$. Another result is that $B\sim I^{-\mu}$  with $\mu$ 
undergoing a jump from 1/3 to 1 at $I_2\approx 60\ mA$. Such a current crossover is clearly visible in the insert of 
figure 2 where we plot together $B$ normalized by $I_0^{1/3}$ and $B$ normalized by $I_1$ versus 
I depending upon whether $I$ is $<I_2$ or $>I_2$, respectively. Here, $I_0$ and $I_1$ are two currents evaluated 
on the basis of the BG model, as will be seen below. It may be seen that below $I_2$, the solid line of slope $\sim -1/3$ 
fits the data very well, whereas the dashed line of slope $\sim -1$  fits them best above $I_2$. It therefrom follows that
$B_{1/3}(I)=\kappa(I_0/I)^{1/3}$ and $B_{1}(I)=\kappa I_1/I$ where the labels $1/3$ and $1$ differentiate between two 
fits below and above $I_2$, respectively. In both these equations, $\kappa\approx 0.6$ is a dimensionless constant. 
It should be noted that such a crossover at $I_2$ is clearly observed from $V/I-I$ curves as well (see Fig.\ 1), 
where a sudden increase of $V/I$ suggesting an increasing vortex motion, occurs at $I_2$. Second, we plot in Fig.\ 3 
and its insert the natural logarithm of $V/I$ versus $I^{-1/3}$ for $I<I_2$ and versus $I^{-1}$ for $I>I_2$, respectively. 
We verify that the expression $ln(V/I)=C-D_{\mu}I^{-\mu}$ (expression designated below as $E2$) fits very well our 
experimental data (solid lines) with an exponent $\mu=1/3$ for $I<I_2$ and with $\mu=1$ for $I>I_2$. In $E2$, $D_{\mu}$ 
is labeled using the above convention. The result is that $C$ is a constant (within the experimental errors), 
while $D_{\mu}$ depends on $\theta$. In Fig.\ 4 and its lower insert, we show $D_{1/3}$ and $D_{1}$ versus $\tan{\theta}$, 
respectively. A linear variation in $\tan{\theta}$ is observed consistently with expression $E1$. In upper insert of 
Fig.\ 4, we plot together $D_{1/3}/I_0^{1/3}$ and $D_{1}/I_1$ versus $\tan{\theta}$. 
The result is that the data are superimposable onto a single straight line with slope $\approx-0.6$ consistent with 
the value of $-\kappa$, as may be verified by identifying the expression $E1$ with the expression $E2$.\\
	\indent In conclusion, $V/I=R_0\exp{\left[-\left(\kappa'-\kappa \tan{\theta}\right)\left(I_0/I\right)^{1/3}\right]}$
is clearly observed in our experiment for $I< I_2$ (solid lines in Fig. 1) while 
$V/I=R_0\exp{\left[-\left(\kappa'-\kappa \tan{\theta}\right)\left(I_1/I\right)\right]}$ is valid for data above 
$I_2$ (dashed lines in Fig. 1). In both fitting expressions $\kappa\approx 0.6$ and $\kappa'\approx 0.4$ are two parameters 
that {\it a priori} depend on the temperature and filling factor, $I_0=15$ A and $I_1=0.4$ A are evaluated from BG model 
(see below), and $R_0\equiv e^C\approx 1.1$ $\mu\Omega$ which is five orders of magnitude lower than the normal resistance. 
Therefore, for $\theta<\theta_c$ we have rather strong evidence of two separate (albeit related) vortex creep processes 
peculiar to a BG : the half-loop expansion with $\mu=1$ which is cut off by the crossover at $I_2$ into the VRH process 
with $\mu=1/3$. Moreover, we note that for each filling factor investigated in our experiment ($f=0.13,\ 1/3,\ 2/3$), we 
observe a {\it continuation} of the VRH process evidenced at $\theta=0^{\circ}$ \cite{12}. We therefore argue that the BG 
phase remains stable up to a critical tilting angle $\theta_c$, as predicted by NV \cite{2}.\\
	\indent To test the accuracy of the above view, we estimate from the theory \cite{1,2}, first the characteristic current 
scales, secondly the current crossover, and then the energy barriers in Eq.\ (1). $J_0$ only depends on 
$g(\tilde{\mu})$. Although a form of $g(x)$ is not yet available, an estimation of $g(\tilde{\mu})$ can be done in terms 
of $\gamma$, the bandwidth of pinning energies due to the disorder, and hence, 
$g(\tilde{\mu})\approx1/\left(d^2\gamma\right)$. We take $\gamma=t_d+\gamma_i$ where $\gamma_i$ arise from some on-site 
disorder, and $t_d\approx U/\sqrt{E_k/k_BT}\exp\left( -\sqrt{2}E_k/k_BT\right)$ is due to structural disorder \cite{2}. 
Assuming the variation in the defect diameters as the first cause for random on-site energies, we approximate $\gamma_i$ 
to the width of the distribution of pinning energies $\tilde{P}(U_K)=P(c_K)dc_K/dU_K$ where the pinning energy and the 
defect radius $c_K$ are related to one another through the formula 
$U_K=\epsilon_0\ln{\left(1+\left(c_K/\sqrt{2}\xi_{ab}\right)^2\right)}^{1/2}$ and $P(c_K)$ designates 
a probability density function for the defect radii. In the following estimation of $\gamma_i$ , we use realistic Gaussian 
law centered at $c_0=45$ \AA \ with a standard deviation of $6$ \AA, as shown in figure 1 of Ref.\ \onlinecite{12}. 
Then, taking $E_K=d\sqrt{\tilde{\epsilon}_1U}$ with 
$\tilde{\epsilon}_1\approx\epsilon\epsilon_0\ln\left( a_0/\xi_{ab}\right)$ where $\epsilon$ is the anisotropic parameter 
and $\epsilon_0=\phi_0^2/\left(4\pi\mu_0\lambda_{ab}^2\right)$, and with $U=U_0f(T/T^{\star})$ where 
$T^{\star}=max[c_0,\sqrt{2}\xi_{ab}]\sqrt{\tilde{\epsilon}_1U_0}$ is the energy scale for the pinning 
and $f(x)=x^2/2\exp\left(-2x^2\right)$ accounts for thermal effects, and using appropriate parameters for 
BSCCO ($\lambda_{ab}\approx 1850$ \AA, $\xi_{ab}\approx20$ \AA and $\epsilon\approx1/200$) we obtain 
$J_0\approx\gamma/(\phi_0d)\approx10^9A/m^2$ and $J_1\approx U/(\phi_0d)\approx 2.6.10^7A/m^2$. Assuming uniform currents 
into the sample we have $I_0\approx15$ A and $I_1\approx0.4$ A. The basis for the estimate of the current crossover 
$J_2$ between the VRH process and the half-loops regime corresponds to the strongly dispersive situation with 
$U/\gamma=J_1/J_0\equiv I_1/I_0\approx2.6.10^{-2}\ll 1$, so that 
$I_2\approx(U/\gamma)^{1/2}I_1\approx(U/\gamma)^{3/2}I_0$
with $I_2<I_1<I_0$ \cite{1}, and thus, we obtain $I_2\approx6.10^{-2} A$ in excellent agreement 
with the experiment (see Fig.\ 1). Below and above this current crossover, the energy barriers against 
vortex motion are $\tilde{E}_k(I_0/I)^{1/3}$ and  $\tilde{E}_k(I_1/I)$, respectively, where $\tilde{E}_k$ is given by 
Eq.\ (2). Using again the above usual parameters of BSCCO, we estimate $E_k/k_BT\approx0.5$ and 
$\epsilon^2\phi_0dH_{\perp}/k_BT\approx0.7\tan{\theta}$, which correspond to the values of our fitting parameters 
$\kappa'$ and $\kappa$ (see upper insert in Fig.\ 4). Thus, the experiment 
is also in good agreement with the energy barriers theoretically predicted below and above $I_2$.\\
	\indent Finally, we pass on to the linear regime observed above $\theta_c$ (see Fig.\ 1). For $\theta> \theta_c$, 
Hwa et al. \cite{6} predict on the basis of the CIC transition, the appearance of free moving chains of kinks oriented 
in the $H_{\perp}$ direction leading to a critical behavior of the linear resistivity 
$\rho\sim(\tan{\theta}-\tan{\theta_c})^{\nu}$ characterized by an 
exponent $\nu=1/2$ (or $\nu=3/2$) in (1+1) [or (2+1)] dimensions. This type of behavior with $\nu=1/2$ is evident
 in Fig.\ 5 which shows a plot of $R/R_C$ versus $\tan{\theta}-\tan{\theta_c}$ for three temperatures 
less than $T_{BG}(0)$. 
Here, $R_C$ is a scaling parameter comparable to $R_0$, and the critical tilt angle $\theta_c(T)$ has been 
determined fitting $R=R_C(\tan{\theta}-\tan{\theta_c})^{1/2}$
 to the data. In Fig.\ 2, the arrow indicates $\theta_c$ obtained from this fit. The insert of the figure 5 shows 
$T_{BG}(0)-T_{BG}(\theta)/T_{BG}(0)$ as a function of $\tan{\theta}$ where $T_{BG}(\theta)$ is obtained 
through the inversion of $\theta_c$ and $T$. The solid line shown in insert is a fit to the data 
following $T_{BG}(0)-T_{BG}(\theta)\sim |\tan{\theta}|^{1/\nu_{\perp}}$ with $\nu_{\perp}=1.0\pm 0.1$, as recently 
suggested by Lidmar and Wallin from numerical simulations \cite{14} and observed in $(K,Ba)BiO_3$ \cite{15}. Note that 
such a value of $\nu_{\perp}$ is excellently consistent with the result previously found using the scaling theory of the 
BG transition at $\theta=0^{\circ}$ \cite{12} .\\
	\indent In summary, we have shown via a detailed quantitative analysis of the transport properties 
in BSSCO single crystal with columnar defects that a TME is supported by the experimental 
evidence of the creep processes (VRH and half-loop) showing the stability of the BG phase when
 ${\bf H}$ is tilted away from the column direction. Above the critical angle, the commensurate-incommensurate 
transition have been evidenced. Finally, our results indicate that the vortices behave as well-connected lines 
in accordance with the idea that columnar defects effectively increase interlayer coupling in highly layered 
superconductors \cite{16,17}.\\

  \begin{figure}[F1]
	\centerline{\epsfxsize=7 cm $$\epsfbox{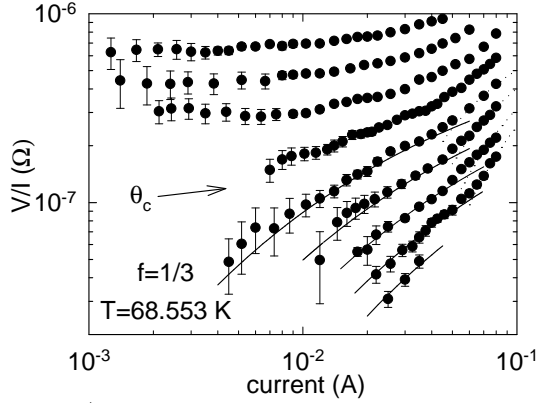}$$}
    \caption{$V/I-I$ curves for tilted magnetic fields in an irradiated $Bi_2Sr_2CaCu_2O_{8+y}$ crystal. 
>From the right to the left $\theta=0,\ 5,\ 10,\ 15,\ 20,\ 25,\ 30,\ 35$ and $40^{\circ}$. 
The curved lines are a fit of the Bose glass theory to the data.
      }
  \end{figure}

  \begin{figure}[F2]
	\centerline{\epsfxsize=7 cm $$\epsfbox{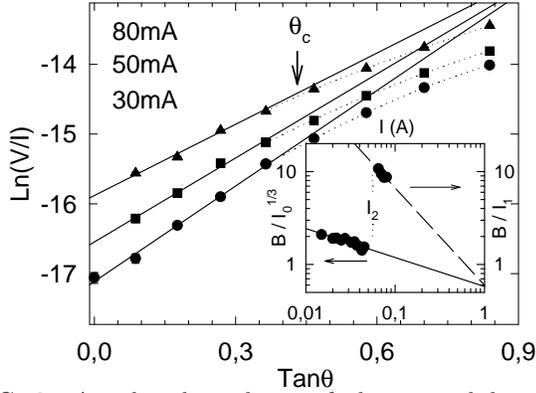}$$}
    \caption{Angular dependence of the natural logaritmic of $V/I$ for different currents. The solid lines are a fit of
$\ln(V/I)=A+B\tan\theta$ to the data; the dotted lines are a guide for the eye. Note that as the critical angle is 
approached from above, the linear resistivity drops continuously to zero (see Fig. 5). Insert: $B/I_0^{1/3}$  vs. I 
(left axis) and $B/I_1$  vs. I (right axis) with $I_0$ and $I_1$ determined from theory. 
The straight lines are least-square fits wherefrom we obtain $\mu=0.31\pm 0.05$  (solid line) for $I<I_2$ and 
$\mu=1.0\pm 0.1$  (dashed line) for $I>I_2$.
      }
  \end{figure}

  \begin{figure}[F3]
	\centerline{\epsfxsize=7 cm $$\epsfbox{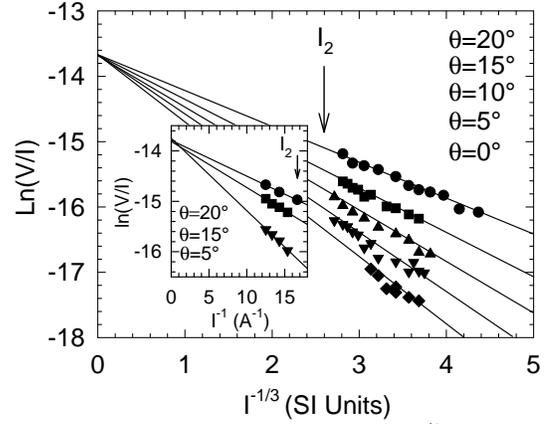}$$}
    \caption{Natural logaritmic of $V/I$ vs. $I^{-1/3}$ for $I<I_2$ and $\theta<\theta_c$. Insert: natural logaritmic of 
$V/I$ vs. $I^{-1}$ for $I>I_2$ and $\theta<\theta_c$. In both plots the lines are a fit to the data using the form  
$\ln(V/I)=C-D_{\mu}I^{-\mu}$ with $\mu=1/3$ or $1$ according to whether $I$ is $<I_2$ or $> I_2$.
      }
  \end{figure}

  \begin{figure}[F4]
	\centerline{\epsfxsize=7 cm $$\epsfbox{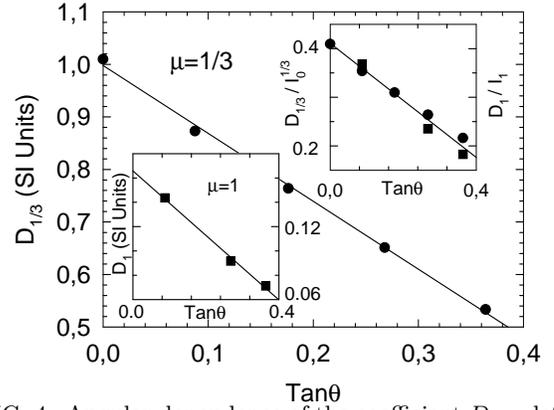}$$}
    \caption{Angular dependence of the coefficient $D_{1/3}$ defined in the regime $I<I_2$ (see Fig. 3). Lower insert: 
angular dependence of the coefficient $D_1$  defined in the regime $I>I_2$. Upper insert: angular dependence of 
$D_{1/3}/I_0^{1/3}$  and $D_1/I_1$ . In each plot the straight line is a least-square fit.
      }
  \end{figure}

  \begin{figure}[F5]
	\centerline{\epsfxsize=7 cm $$\epsfbox{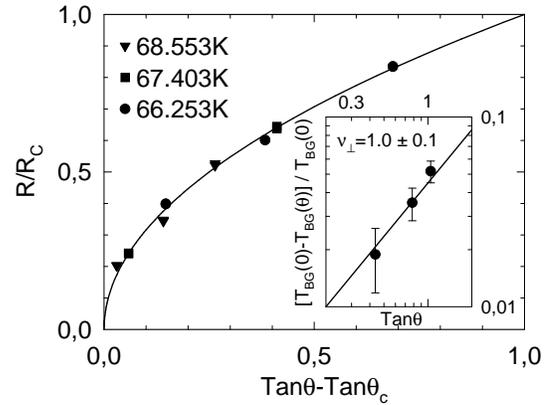}$$}
    \caption{Critical behavior of the linear resistance at different temperatures for $\theta>\theta_c(T)$. 
The curved line is a fit to the data using $R=R_c\left(\tan\theta-\tan\theta_c\right)^{1/2}$
  in accordance with a CIC transition scenario. Insert: Angular dependence of the Bose-glass transition. 
The solid line is a fit of $T_{BG}(0)-T_{BG}(\theta)\sim |\tan\theta|^{1/\nu_{\perp}}$  with $\nu_{\perp}=1.0\pm 0.1$ 
to the data.
      }
  \end{figure}

\end{multicols}
\end{document}